\documentclass[conference,usenatbib]{basi}
\usepackage[T1]{fontenc}
\usepackage[british]{babel}
\usepackage[varg]{txfonts}
\usepackage{rotating}
\usepackage{dcolumn}

\usepackage{aas_macros}
\usepackage{epsfig}

\newcommand{\bei}{\begin{itemize}}
\newcommand{\eei}{\end{itemize}}
\newcommand{\bef}{\begin{figure}}
\newcommand{\eef}{\end{figure}}
\newcommand{\ben}{\begin{enumerate}}
\newcommand{\een}{\end{enumerate}}
\newcommand{\beq}{\begin{equation}}
\newcommand{\eeq}{\end{equation}}
\newcommand{\ber}{\begin{eqnarray}}
\newcommand{\eer}{\end{eqnarray}}

\newcommand{\bb}{\bf B}

\newcommand{\pa}{\partial}

\newcommand{\vb}{\bf v}

\newcommand{\gcc}{\mbox{${\rm g} \, {\rm cm}^{-3}$}}
\newcommand{\mdot}{\mbox{$\dot{\rm M}$}}
\newcommand{\msun}{\mbox{{\rm M}$_{\odot}$}}

\newcommand{\gsim}{\raisebox{-0.3ex}{\mbox{$\stackrel{>}{_\sim} \,$}}}

\begin{document}
\title[Magnetic Fields of Neutron Stars]{\sl{Magnetic Fields of Neutron Stars : The AMXP Connection}}
\author[Konar]%
       {Sushan Konar\thanks{email: \texttt{sushan@ncra.tifr.res.in}},
       NCRA-TIFR, Pune University Campus, Pune, India}

\pubyear{2013}
\volume{**}
\pagerange{**--**}

\date{Received --- ; accepted ---}

\maketitle
\label{firstpage}

\begin{abstract}
This  article  briefly  reviews  our current  understanding  (or  lack
thereof) of the evolution of magnetic fields in neutron stars, with an
emphasis on the binary systems. In particular, the significance of the
newly emerging  population of accreting millisecond  pulsars (AMXP) is
discussed.
\end{abstract}
%

\section{The Neutron Star Menagerie}
\label{ns}

In recent  years we  have been confronted  by a variety  of apparently
disparate  observational classes of  neutron stars.   Fortunately, the
processes responsible  for energy  generation in these  2000+ objects,
even  though their  radiation span  almost the  entire electromagnetic
spectrum,  belong basically  to three  categories.   Consequently, the
neutron stars can be classified  into three types in accordance to the
manner of energy generation.

{\bf A.  Rotation Powered :}  The classical radio pulsars ({\bf PSR}),
these are  powered by  the loss of  rotational energy due  to magnetic
braking. Millisecond radio pulsars ({\bf MSRP}) are simply a sub-class
of RPPs, albeit being very bright in $\gamma$-rays.

{\bf  B.   Accretion  Powered  :}  Accreting neutron  stars  in  HMXBs
typically  show up  as high-magnetic  field  accretion-powered pulsars
({\bf APP}).  Whereas neutron stars in LMXBs have weak magnetic fields
and  the emission  is usually  not  pulsed. However,  in systems  with
extremely low rates of mass transfer  the neutron stars may show up as
bursting X-ray transients ({\bf XRT}) or the dramatic {\bf AMXP}s.

{\bf C. Internal Energy Powered  :} A heterogeneous class with objects
powered  by some  form of  internal  energy. The  {\bf magnetars}  are
thought  to shine  due to  the  decay of  their super-strong  magnetic
fields.  The  soft gamma-ray repeaters  ({\bf SGR}) and  the anomalous
X-ray  pulsars  ({\bf AXP})  are  most  likely different  evolutionary
phases  of magnetars  themselves.  The X-ray  bright isolated  neutron
stars ({\bf INS}) and the central compact objects ({\bf CCO}) are most
likely powered by their residual  thermal energy and/or the decay of a
strong  magnetic field.  The rotating  radio transients  ({\bf RRAT}),
characterised  by  their   powerful  single-pulse  radio  bursts,  are
suspected to be extreme  cases of nulling/intermittent pulsars. Though
the  nulling pulsars  are powered  by rotation,  the energy  source of
RRATs is not likely to be the same.

The  challenge of  neutron star  research has  always been  to  find a
unifying theme to explain this menagerie.  The magnetic field, ranging
from $10^8$~G in  MSRPs to $10^{15}$~G in magnetars,  has been central
to this theme. It plays an important role in determining the evolution
of  the  spin, the  radiative  properties  and  the interaction  of  a
neutrons  star with  its  surrounding medium.   The  evolution of  the
magnetic field is therefore vital to our knowledge of the neutron star
physics as a whole.

\bef
\centerline{\includegraphics[width=11.0cm]{konar_01.ps}}
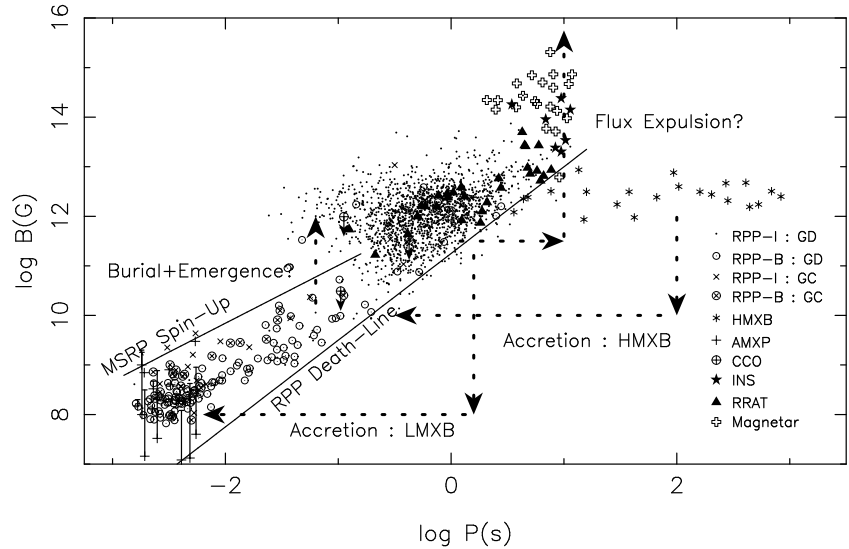
\caption{The neutron star menagerie in the $P-B$ plane and possible
evolutionary pathways. The data have been obtained from a number
of publicly available resources (see \cite{bhatt13} for detailed 
references). The range in AMXP field values are due to difference in
measurements obtained using different techniques~\citep{mukhe13b}. }
\eef

\section{Evolution of the Magnetic Field}

The magnetic field  in a neutron star either  evolves spontaneously or
as  a consequence of  material accretion.  It has  been argued  that a
physical  model of  field evolution  should satisfy  the observational
constraints that  relatively little  magnetic field decay  should take
place in  isolated radio pulsar population, while  accretion should be
able  to  reduce the  surface  field  strength  by several  orders  of
magnitude~\citep{bhatt02}.  However,  in modification to  this original
scenario,  a theory  of  magneto-thermal evolution  has recently  been
developed to  understand the evolutionary links  between the different
types of isolated neutron stars~\citep{pons09}.

There is no  consensus regarding the generation of  the magnetic field
in neutron stars. The field could  be - {\bf a.} a fossil remnant from
the  progenitor star in  the form  of Abrikosov  fluxoids of  the core
proton  superconductor~\citep{baym69a,ruder72}; {\bf b.}   - generated
by   the  turbulent  currents   inside  the   core  before   it  turns
superconducting~\citep{thomp01};  or {\bf c.}   - entirely  confined to
the    solid    crust    generated    therein    by    thermo-magnetic
currents~\citep{bland83,urpin86}.

However, in  most cases the processes responsible  for field evolution
(for example ohmic dissipation, ambipolar diffusion or Hall drift) can
only be effective if the currents supporting the field are located (or
are relocated in the course  of evolution) in the crustal region which
has  metal-like  transport  properties.   The simplest  and  the  only
mechanism resulting in  a permanent decrease of the  field strength is
ohmic dissipation of the  currents.  It also successfully explains the
MSRP  generation via  recycling  of ordinary  radio  pulsars in  X-ray
binaries,  where the  magnetic field  decreases by  several  orders of
magnitude in accretion heated crust.

The basic  physics underlying the model  of magneto-thermal evolution,
invoked for the  isolated neutron stars, is also  essentially the same.
On the  one hand the field  evolution is sensitively  dependent on the
transport  properties  (namely  the  electrical conductivity)  of  the
crust.   Since  conductivity is  a  function  of temperature,  thermal
evolution  affects field  evolution.   On the  other  hand, the  ohmic
dissipation of the field  generates heat, modifying thermal evolution,
reducing conductivity and affecting  an even faster dissipation of the
magnetic field.

\subsection{Isolated Neutron Stars}

A  new  scenario  is  emerging  out  of  recent  observations  linking
different types  of isolated neutron  stars (see \cite{kaspi10}  for a
detailed review).   Notice that there  is a clear overlap  between the
high magnetic field  ($B > 4 \times 10^{13}$~G)  radio pulsars and the
magnetars  in the  $B-P$  diagram (Fig.[1]).  The magnetar-like  X-ray
burst  exhibited by PSR  J1846-0258 ($  B =  4 \times  10^{13}$~G) has
reinforced  the suggestion  that  such high  field  radio pulsars  are
quiescent   magnetars.   Conversely,  it   has  been   suggested  that
hyper-critical fallback accretion may bury the field to deeper crustal
layers  thereby reducing  the  surface  field, as  seen  in the  CCOs.
Subsequent re-emergence of this buried  field could transform a CCO to
an ordinary radio  pulsar or even to a  magnetar.  Therefore different
combinations  of initial  spin-period, magnetic  field  and submersion
depth  of  the field  may  very well  decide  whether  a neutron  star
manifests  itself  as  an  ordinary  radio pulsar,  a  magnetar  or  a
CCO~\citep{vigan12}.   Similarly,  INSs are  observed  only in  X-ray,
despite being isolated objects.  It is possible that they are actually
similar to  the RPPs and are not  seen as radio pulsars  simply due to
the  misalignment of  emission cones  with  our lines  of sight.   The
neutron stars with strong magnetic  fields are expected to remain at a
relatively  higher temperature due  to field  decay.  This  could then
explain the high (compared to ordinary radio pulsars) X-ray luminosity
of the  INSs. Finally, it has  been argued that  the anomalous braking
index of PSR  J1734-3333 signifies an increase in  its dipolar surface
magnetic  field.  This  is  likely driven  by  the emergence  (perhaps
glitch-induced)  of a  stronger field  buried underneath  the surface,
with  timescales depending  on submersion  depth~\citep{espin11c}.  If
correct,  this process  may chart  a pathway  for the  transition from
ordinary  radio  pulsars  to  magnetars.  It  appears  that  different
flavors of the  isolated neutron stars could, in  fact, be intricately
connected through various evolutionary pathways.

\subsection{Binary Neutron Stars}

Three major physical models have been invoked to explain the evolution
of  the magnetic  field in  binary neutron  stars, namely  -  {\bf a.}
diamagnetic     screening     of     the     field     by     accreted
plasma~\citep{cummi01,choud02,konar04b,payne04,payne07},    {\bf   b.}
spindown-induced                                                   flux
expulsion~\citep{bhatt96c,konar99b,konen00,konen01a}  to  the  crustal
regions, and  {\bf c.}  rapid ohmic  decay of the crustal  field in an
accretion    heated    crust~\citep{geppe94,konar97a,konar99a}.    The
investigation into the consequences of diamagnetic screening have only
recently  begun in  the earnest  and we  exclude it  from  the present
discussion.

\bef
\begin{center}
\includegraphics[angle=-90,width=7.5cm]{konar_02.ps}
\end{center}
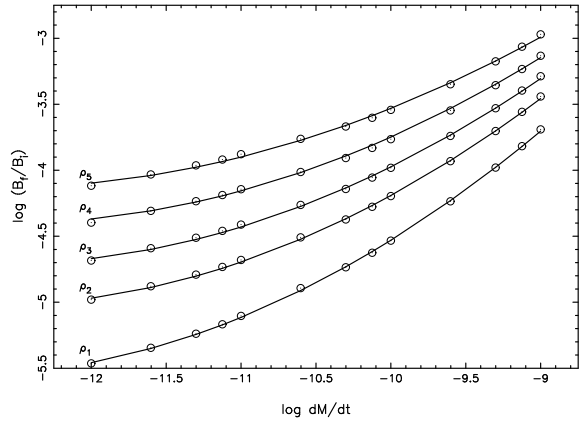
\caption{Final  surface field  as  a function  of  \mdot~ \&  $\rho_c$
  (increasing from $\rho_1$ to $\rho_5$).}
\eef

The other  two models invoke  ohmic decay of  the current loops  for a
permanent decrease  in the field  strength which happens  according to
the induction equation, given by -
\beq
\frac{\pa \bb}{\pa t} 
= \nabla \times (\vb \times \bb) 
  - \frac{c^2}{4\pi} \nabla \times (\frac{1}{\sigma} \nabla \times \bb).
\eeq
In  the  deeper  layers of  the  crust  the  field decay  is  governed
essentially by the electrical  conductivity $\sigma$, and the radially
inward  material  velocity $\vb  \,  (\propto  \mdot/r^2$).  In  turn,
$\sigma$ is  dependent on $\rho_c$,  the density at which  the current
carrying  layers are concentrated,  the impurity  content $Q$  and the
temperature of  the crust $T_c$  (again decided by the  mass accretion
rate \mdot).

Accretion-induced heating reduces  $\sigma$ and consequently the ohmic
decay  time-scale inducing  a  faster  decay.  At  the  same time  the
material movement,  caused by the deposition  of matter on  top of the
crust,  pushes the original  current carrying  layers into  deeper and
denser  regions where the  higher conductivity  slows the  decay down.
Ultimately  the decay  stops  altogether when  the  original crust  is
assimilated      into     regions     with      effectively   infinite
conductivity~\citep{konar97a}.

Therefore,  the final saturation  field of  an accreting  neutron star
depends entirely upon the initial magnetic field, the initial $\rho_c$
and $\mdot$  which determines both $T_c$  and {\vb} as can  be seen in
the  left  panel  of  Fig.[2].  As  is  evident, the  model  of  ohmic
dissipation  is  excellent  in  terms  of  producing  magnetic  fields
observed  in  typical  MSRPs,  starting  from  ordinary  pulsar  field
strengths.

\bef
\begin{center}
\includegraphics[angle=-90,width=7.5cm]{konar_03.ps}
\end{center}
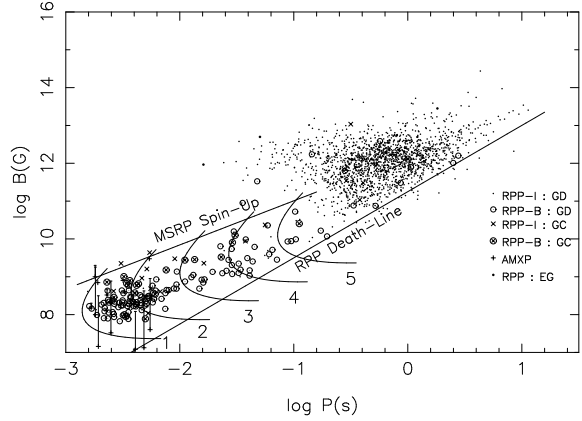
\caption{Final   surface   field   and   spin-period   with   currents
  concentrated  at $\rho_c  =  10^{13}$~\gcc.  Curves  1  to 5  denote
  different     initial     field     strengths     ($10^{11.5}$~G     -
  $10^{13.5}$~G).}
\eef

{\bf  Location of the  magnetic field  :} The  intrinsic uncertainties
associated with the  model of ohmic dissipation are  - a) the impurity
content  of  the crust,  and  b) the  exact  location  of the  current
carrying  layers.   Fortunately,  when  the crustal  temperatures  are
sufficiently high (as realised  in accreting neutron stars) the effect
of impurities  can be entirely  neglected. Consequently, we  find that
the ohmic dissipation  model can be used to  constrain the location of
the current carrying layers inside  a neutron star.  We assume that an
ordinary  neutron  star is  born  with  a  typical magnetic  field  of
$10^{11.5}  - 10^{13.5}$~G.   We find  that in  order to  generate the
observed  population of  MSRPs, the  original current  carrying layers
need to be concentrated $\rho_c \gsim 10^{13}$~\gcc as is shown in the
right panel of Fig.[3]~\citep{konar13b}.

\section{The AMXP-MSRP Connection}

It has long been understood that the neutron stars are spun up by mass
transfer from a stellar companion  in an LMXB and are thereby recycled
to  MSRPs,  with an  attendant  reduction  in  the magnetic  field  as
discussed in  the previous section.   The April 1998 detection  of the
first AMXP (SAX J1808.4-3658) provided the first direct proof of this
model~\citep{wijna98}.  

\begin{table}
\centering
 \begin{tabular}{|l|r|r|r|} \hline
 & $P_{\rm spin}^{\rm av}$ (ms) 
 & $B^{\rm av}$ ($10^8$~G)
 & $P_{\rm orb}^{\rm av}$ (hr) \\ 
 MSRP  ( $P_s \leq 30$~ms) :   &&& \\  
 isolated & 5.16 (19) & 3.45 (19) & \\
 binary   & 5.32 (65) & 2.51 (65) & 768 (65) \\
&&& \\
 AMXP :  
          & 3.81 (30) & 4.93 (14) & 0.16 (24)\\ \hline
 \end{tabular}
\caption[]{Average surface magnetic field, spin and orbital period
of AMXPs and MSRPs. The number of objects in a particular group is
within brackets.}
\end{table}

\bef
\centerline{\includegraphics[angle=-90,width=7.5cm]{konar_04.ps}}
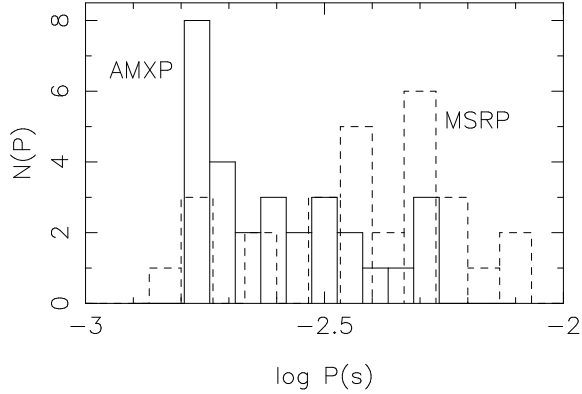
\caption{The $P_{\rm spin}$ histogram for AMXPs and isolated MSRPs.}
\eef

The population  of AMXPs  has been growing  rapidly over the  last few
years, taking the count to 30  (inclusive of standard AMXPs as well as
millisecond   bursting    sources)~\citep{bhatt13}.    These   objects
typically belong to ultra-compact binaries undergoing mass transfer at
very low  rates from low-mass companions.  Though  the average $P^{\rm
  spin}$ of  the AMXPs is smaller  than that of  the millisecond radio
pulsars  (MSRP), the average  $B^{\rm surface}$  tends to  be slightly
higher in AMXPs (as seen in the above table).  It is likely that there
exist possible selection effect for  high-B objects.  And we also need
to concede that the uncertainties  in the estimates made for the field
strength are also rather large.

According to \cite{bilds01} objects  like SAX-J1808 are progenitors of
fast MSRPs with very short orbital periods which have undergone a very
long  period  (Gyr)  of  accretion  at very  low  rates  (\mdot  $\sim
10^{-11}$~\msun/yr).  However,  there appears to be  a generic problem
associated with  the end-products of the  AMXPs.  It can  be seen from
table(1) that  the average  orbital period of  the AMXPs is  very much
smaller than  that of the  MSRPs. It is  true that there exist  a bias
against detecting MSRPs with  very small orbital periods in comparison
to AMXPs. But given that many  of the AMXPs (including SAX J1808) show
{\em black-widow}  traits it  is suggested that  most of  the observed
AMXPs would  end up as isolated  MSRPs.  However, a  comparison of the
spin-period  distribution of AMXPs  and the  isolated MSRPs  show that
they   are   completely   mismatched    (as   seen   in   the   figure
above)~\citep{konar13b}.

We  conclude that  even though  the  observed population  of AMXPs  is
consistent  with pure ohmic  dissipation model,  it however,  does not
really  mimic the MSRP  population. This  leaves us  with a  couple of
puzzles regrading  the nature of -  {\bf 1.} the real  end products of
the  observed AMXPs,  and {\bf  2.}  the  progenitor population  of the
observed MSRPs. It is worth noting that investigations into the nature
of binary  evolution also  suggest that different  types of  LMXBs may
produce different kinds of MSRPs~\citep{chen13}.


\bibliography{adsrefs}

\begin{thebibliography}{}

\bibitem[\protect\citeauthoryear{{Baym}, {Pethick}, \& {Pines}}{{Baym}
  et~al.}{1969}]{baym69a}
{Baym} G., {Pethick} C.,  {Pines} D., 1969, \nat, 224, 673

\bibitem[\protect\citeauthoryear{{Bhattacharya}}{{Bhattacharya}}{2002}]{bhatt0%
2}
{Bhattacharya} D., 2002, Journal of Astrophysics and Astronomy, 23, 67

\bibitem[\protect\citeauthoryear{{Bhattacharya} \& {Datta}}{{Bhattacharya} \&
  {Datta}}{1996}]{bhatt96c}
{Bhattacharya} D.,  {Datta} B., 1996, \mnras, 282, 1059

\bibitem[\protect\citeauthoryear{{Bhattacharya}, {Konar}, \&
  {Mukherjee}}{{Bhattacharya} et~al.}{2013}]{bhatt13}
{Bhattacharya} D., {Konar} S.,  {Mukherjee} D., 2013, {\em in prep.}

\bibitem[\protect\citeauthoryear{{Bildsten} \& {Chakrabarty}}{{Bildsten} \&
  {Chakrabarty}}{2001}]{bilds01}
{Bildsten} L.,  {Chakrabarty} D., 2001, \apj, 557, 292

\bibitem[\protect\citeauthoryear{{Blandford}, {Applegate}, \&
  {Hernquist}}{{Blandford} et~al.}{1983}]{bland83}
{Blandford} R.~D., {Applegate} J.~H.,  {Hernquist} L., 1983, \mnras, 204, 1025

\bibitem[\protect\citeauthoryear{{Chen} et~al.}{{Chen} et~al.}{2013}]{chen13}
{Chen} H.-L., {Chen} X., {Tauris} T.~M.,  {Han} Z., 2013, ArXiv e-prints

\bibitem[\protect\citeauthoryear{{Choudhuri} \& {Konar}}{{Choudhuri} \&
  {Konar}}{2002}]{choud02}
{Choudhuri} A.~R.,  {Konar} S., 2002, \mnras, 332, 933

\bibitem[\protect\citeauthoryear{{Cumming}, {Zweibel}, \& {Bildsten}}{{Cumming}
  et~al.}{2001}]{cummi01}
{Cumming} A., {Zweibel} E.,  {Bildsten} L., 2001, \apj, 557, 958

\bibitem[\protect\citeauthoryear{{Espinoza} et~al.}{{Espinoza}
  et~al.}{2011}]{espin11c}
{Espinoza} C.~M., {Lyne} A.~G., {Kramer} M., {Manchester} R.~N.,  {Kaspi}
  V.~M., 2011, \apjl, 741, L13

\bibitem[\protect\citeauthoryear{{Geppert} \& {Urpin}}{{Geppert} \&
  {Urpin}}{1994}]{geppe94}
{Geppert} U.,  {Urpin} V., 1994, \mnras, 271, 490

\bibitem[\protect\citeauthoryear{{Kaspi}}{{Kaspi}}{2010}]{kaspi10}
{Kaspi} V.~M., 2010, Proceedings of the National Academy of Science, 107, 7147

\bibitem[\protect\citeauthoryear{{Konar} \& {Bhattacharya}}{{Konar} \&
  {Bhattacharya}}{1997}]{konar97a}
{Konar} S.,  {Bhattacharya} D., 1997, \mnras, 284, 311

\bibitem[\protect\citeauthoryear{{Konar} \& {Bhattacharya}}{{Konar} \&
  {Bhattacharya}}{1999a}]{konar99a}
{Konar} S.,  {Bhattacharya} D., 1999a, \mnras, 303, 588

\bibitem[\protect\citeauthoryear{{Konar} \& {Bhattacharya}}{{Konar} \&
  {Bhattacharya}}{1999b}]{konar99b}
{Konar} S.,  {Bhattacharya} D., 1999b, \mnras, 308, 795

\bibitem[\protect\citeauthoryear{{Konar} \& {Choudhuri}}{{Konar} \&
  {Choudhuri}}{2004}]{konar04b}
{Konar} S.,  {Choudhuri} A.~R., 2004, \mnras, 348, 661

\bibitem[\protect\citeauthoryear{{Konar}, {Mukherjee}, \&
  {Bhattacharya}}{{Konar} et~al.}{2013}]{konar13b}
{Konar} S., {Mukherjee} D.,  {Bhattacharya} D., 2013, {\em in prep.}

\bibitem[\protect\citeauthoryear{{Konenkov} \& {Geppert}}{{Konenkov} \&
  {Geppert}}{2000}]{konen00}
{Konenkov} D.,  {Geppert} U., 2000, \mnras, 313, 66

\bibitem[\protect\citeauthoryear{{Konenkov} \& {Geppert}}{{Konenkov} \&
  {Geppert}}{2001}]{konen01a}
{Konenkov} D.~Y.,  {Geppert} U., 2001, Astronomy Letters, 27, 163

\bibitem[\protect\citeauthoryear{{Mukherjee} et~al.}{{Mukherjee}
  et~al.}{2013}]{mukhe13b}
{Mukherjee} D., {Bult} P., {van der Klis} M.,  {Bhattacharya} D., 2013, {\em in
  prep.}

\bibitem[\protect\citeauthoryear{{Payne} \& {Melatos}}{{Payne} \&
  {Melatos}}{2004}]{payne04}
{Payne} D.~J.~B.,  {Melatos} A., 2004, \mnras, 351, 569

\bibitem[\protect\citeauthoryear{{Payne} \& {Melatos}}{{Payne} \&
  {Melatos}}{2007}]{payne07}
{Payne} D.~J.~B.,  {Melatos} A., 2007, \mnras, 376, 609

\bibitem[\protect\citeauthoryear{{Pons}, {Miralles}, \& {Geppert}}{{Pons}
  et~al.}{2009}]{pons09}
{Pons} J.~A., {Miralles} J.~A.,  {Geppert} U., 2009, \aap, 496, 207

\bibitem[\protect\citeauthoryear{{Ruderman}}{{Ruderman}}{1972}]{ruder72}
{Ruderman} M., 1972, \araa, 10, 427

\bibitem[\protect\citeauthoryear{{Thompson} \& {Murray}}{{Thompson} \&
  {Murray}}{2001}]{thomp01}
{Thompson} C.,  {Murray} N., 2001, \apj, 560, 339

\bibitem[\protect\citeauthoryear{{Urpin}, {Levshakov}, \& {Iakovlev}}{{Urpin}
  et~al.}{1986}]{urpin86}
{Urpin} V.~A., {Levshakov} S.~A.,  {Iakovlev} D.~G., 1986, \mnras, 219, 703

\bibitem[\protect\citeauthoryear{{Vigan{\`o}} \& {Pons}}{{Vigan{\`o}} \&
  {Pons}}{2012}]{vigan12}
{Vigan{\`o}} D.,  {Pons} J.~A., 2012, \mnras, 425, 2487

\bibitem[\protect\citeauthoryear{{Wijnands} \& {van der Klis}}{{Wijnands} \&
  {van der Klis}}{1998}]{wijna98}
{Wijnands} R.,  {van der Klis} M., 1998, \nat, 394, 344

\end{thebibliography}
\bibliographystyle{mnras}

\label{lastpage}
\end{document}